\documentclass[conference, compsoc]{IEEEtran}

\usepackage[pdftex]{graphicx}
\DeclareGraphicsExtensions{.pdf,.jpeg,.png}
\graphicspath{{}{pdf/}}

\usepackage[cmex10]{amsmath}
\usepackage{fixltx2e}
\usepackage[latin1]{inputenc}
\usepackage[english]{babel}
\usepackage{amssymb}
\usepackage{bm}
\usepackage{natbib}

\newcommand{\eq}[1]{Eq.\,(\ref{#1})}

\renewcommand{\vec}[1]{\mathbf{#1}}
\newcommand{\fig}[1]{Fig.\,\ref{#1}}
\newcommand{\tab}[1]{Tab.\,\ref{#1}}
\newcommand{\ba}[1]{\begin{array}{#1}}
\newcommand{\ea}{\end{array}}
\newcommand{\etal}{~{\it et\,\,al.\,}}
\newlength{\figwidth}

\renewcommand{\baselinestretch}{1.0}
\renewcommand{\theta}{\vartheta}

\begin{document}
\setlength{\figwidth}{0.7\linewidth}

\title{A Lattice Boltzmann study of flow along patterned surfaces and through channels with alternating slip length}

\author{\IEEEauthorblockN{Nayaz Khalid Ahmed}
\IEEEauthorblockA{Department of Chemical Engineering\\
National Institute of Technology\\ 
Tiruchirappalli, India 620 015\\
khalid.nitt\@yahoo.com}
\and
\IEEEauthorblockN{Martin Hecht}
\IEEEauthorblockA{Institute for Computational Physics, \\ 
University of Stuttgart\\
70569 Stuttgart, Germany\\\
hecht\@icp.uni-stuttgart.de}
}

\maketitle

\begin{abstract}
\boldmath In microfluidics, varying wetting properties, expressed in terms 
of the local slip length, can be used to influence the flow of
a liquid through a device. We study flow past surfaces
on which the slip length is modulated in stripes. We find that the effective slip length for such a flow can be expressed as a function of the individual slip lengths on the stripes. The angle dependence of the effective slip is in excellent agreement with a recent theory describing the slip length as a 
tensorial quantity. This tensorial nature allows to induce a transverse flow, which can be used in micro mixers to drive a vortex. In our simulations of a flow through a square channel with patterned surfaces we see a homogeneous rotation about the direction of the flow. We investigate the influence of patterns of cosine shaped varying local slip on the flow field depending on the orientation of the pattern and find the largest effective slip length for periods of stripes parallel to the flow direction. 
\end{abstract}

\section{Introduction}
  
Microdevices tend to behave very differently from devices that are used in everyday life. Surface tension effects are dominant at these scales, and micropumps and microvalves have been fabricated taking advantage of this principle\,\cite{Beskok05}. Inherent in these technologies is the need to develop the fundamental science of small devices. 

To understand the fluid dynamics in microflows, the classical Navier-Stokes equations cannot be applied due to the breakdown of the continuum assumption. Recent computer simulations apply molecular dynamics\,\cite{Rapaport95}. However, these simulations are usually limited to some tens of thousands of particles, length scales of nanometers and time scales of nanoseconds\,\cite{lauga-brenner-stone}. To overcome these shortcomings, mesoscopic simulation methods such as the Lattice Boltzmann method (LBM) are more applicable and have been recognized as a promising approach for simulation of microflows\,\cite{chen-doolen98}. It has been shown that in the macroscopic limit, the Navier Stokes equations can be recovered\,\cite{chen-chen-matthaeus,Higuera89}. However, most previous LBE models virtually correspond to the Navier-Stokes equations on the length scale of the lattice constant, and when these models are applied to near-continuum flows, the slip boundary conditions must be perfectly specified to capture possible boundary effects on the flow behavior. Thus for the simulation of fluid flows with the LBE method, the development of slip-flow boundary conditions has been a critical issue.

In this paper, we use a recently proposed velocity boundary condition for the LBE that is independent of the relaxation process during collision and contains no artificial 
slip\,\cite{Ahmed09a} and thus simulate slip flows by the use of a parameter $\zeta$ that defines the surface tendency to cause slip. Thus we study flow past surfaces on which the slip length is modulated in stripes.


\section{Simulation Method}
\label{sec_method}

\subsection{Lattice Boltzmann Method} The lattice Boltzmann method (LBM) is a numerical method to solve the Boltzmann equation\, \eq{eq_boltzmann} on a discrete lattice\,\cite{chen-doolen98}. The Boltzmann equation expresses how the probability $f(\vec{x},\vec{v},t)$
of finding a particle with velocity $\vec{v}$ at a position $\vec{x}$ and at 
time $t$ evolves with time:
\begin{equation}
 \label{eq_boltzmann}
  \frac{{\rm d}f}{{\rm d}t} =
  \vec{v}\cdot\nabla_\vec{x}f + 
  \vec{F}\cdot\nabla_\vec{p}f +
  \frac{\partial f}{\partial t} =
  \hat\Omega(f)\,,
\end{equation}
where $\vec{F}$ denotes an external body force, $\nabla_{\vec{x},\vec{p}}$ the
gradient in position and momentum space, and 
$\hat\Omega(f)$ denotes the BGK collision-operator\,\cite{BGK}. 

Our simulation takes place on a 3-dimensional lattice. We use a D3Q19 lattice with the vectors $\vec{c}_i$ pointing to the various neighboring sites according to the index $i$ as explained in our previous work\,\cite{Ahmed09a}. 
Note that we express all quantities in lattice units, i.e., time is measured 
in units of update intervals and length is measured in units of the lattice constant. 
 

\subsection{Boundary Conditions}

On the wall boundary nodes, the distribution function assigned to vectors $\vec{c}_i$ pointing out of the lattice move out of the computational domain in the propagation step, and the ones assigned to the opposing vectors are undetermined because there are no nodes which the distributions could come from. Therefore, on the boundary nodes, special rules have to be applied.

Following\,\cite{Succi01}, we interpolate by means of linear combination between the above two boundary conditions to study slip flows\cite{Ahmed09a}. The respective weight of the no-slip contribution and the full slip contribution is controlled through a slip parameter $\zeta$ by the following general relation:
\begin{equation}
\label{eq_slip_param}
f_i^{ps} = \zeta \* f_i^{fs} + (1-\zeta) \* f_i^{ns} \,,
\end{equation}
where $f_i^{ps}$ denotes the density probability function for flow with the partial-slip condition, $f_i^{ns}$ denotes the density probability function of the no-slip boundary condition, and $f_i^{fs}$ stands for the specular reflection boundary condition from which full-slip condition may be obtained. 

For the no-slip boundary condition, we use the formulation presented recently by Hecht and Harting\,\cite{Hecht2009}. This boundary condition is an explicit local on-site second order flux boundary condition for 3D LB simulations on a D3Q19 lattice, which is independent of the relaxation process during collision and contains no artificial slip. For full-slip boundary condition, specular reflection, as suggested in Ref.\,\cite{Succi01}, is used which implies no tangential momentum transfer to the wall.

The slip parameter is defined such that at a value of zero, the no-slip boundary condition is applied, while at a value of one, the full-slip boundary condition is applied. Thus, slip flows can be easily simulated by adjusting the value of $\zeta$ between zero and one according to physical parameters such as the roughness and hydrophobicity of the surface.


In our previous work\,\cite{Ahmed09a} we have shown that the partial slip boundary condition \eq{eq_slip_param} shows a slip length which is independent of the shear rate. The slip length, measured in lattice units and obtained at a given value of the slip parameter, is independent of the channel width and of the density of the fluid.
It depends linearly on the BGK relaxation time $\tau$ of the LB simulation.


By means of the slip parameter  $\zeta$ the slip length $b$ can be adjusted in our simulation. The slip length measured in units of the lattice constant $a$ is given by 
\begin{equation}
\frac{b}{a} = \frac{\tau\,\zeta}{3\,(1 - \zeta)}\,,
\label{eq_sliplen}
\end{equation}
which matches the numerical results with a relative error of less than  $0.03\%$, 
or $10^{-3}$ lattice units, on the entire range of $\zeta$. 
The slip diverges for $\zeta \rightarrow 1$ because this is the full slip case, and it approaches zero in the no-slip case for $\zeta \rightarrow 0$. 

\section{Results and Discussion}

\paragraph*{Simulation Setup:} The simulation volume consists of a cubic box of 32 nodes in the $x$, $y$, and $z$-directions. Periodic boundary conditions are set up along the $y$ and $z$-directions. 
We use the BGK relaxation time $\tau=1$ throughout the paper.
The boundary conditions given in the previous section are applied in $x$-direction. A constant accelerating force is applied to the whole domain. The $x$-component of the force is always zero, but the force can be turned continuously in the $yz$-plane. The setup is relaxed up to 70,000 lattice updates until a steady state was reached. The profile of fluid flow between two plates is studied in three dimensions, where the slip parameter $\zeta$ is a function of the 
position on the surface, which reflects the pattern as described below.


\subsection{Textured Walls with alternating Stripes}
In our simulations we study the flow past patterned surfaces. First, we restrict 
ourselves to the case of stripes of equal width with alternating slip parameters
$\zeta_1$ and $\zeta_2$. 
The effective slip length of the flow through such a channel is found to be a third order function of the slip length obtained from homogeneous walls, given by the form:

\begin{eqnarray}
b_{m} &=& A \cdot (b_{1} + b_{2}) + B \cdot (b_{1}^2 + b_{2}^2) \nonumber\\
      &+& C \cdot b_{1} \cdot b_{2} + D \cdot (b_{1}^2 \cdot b_{2} + b_{2}^2 \cdot b_{1}) \label{slip_mod} \\ &+& E \cdot (b_{1}^3 + b_{2}^3) \nonumber
\end{eqnarray}

where $b_m$ is the slip length of the modified wall with stripes, $b_{1}$ and $b_{2}$ are the slip lengths obtained for the homogeneous walls of the same slip parameters $\zeta_1$ and $\zeta_2$, respectively, from \eq{eq_sliplen}.  $A$, $B$, $C$, $D$, and $E$ are prefactors, which do not depend on the slip parameters. However, they still depend on the orientation of the stripes on the wall and on the width of the stripes of which the pattern is formed. Note, that \eq{slip_mod} is chosen to be symmetric in $b_{1}$ and $b_{2}$ which reflects the fact that our stripes all have equal width.

\begin{figure*}[t]
\centering
\begin{minipage}{\linewidth}
\mbox{a)\includegraphics[width=0.47\linewidth]{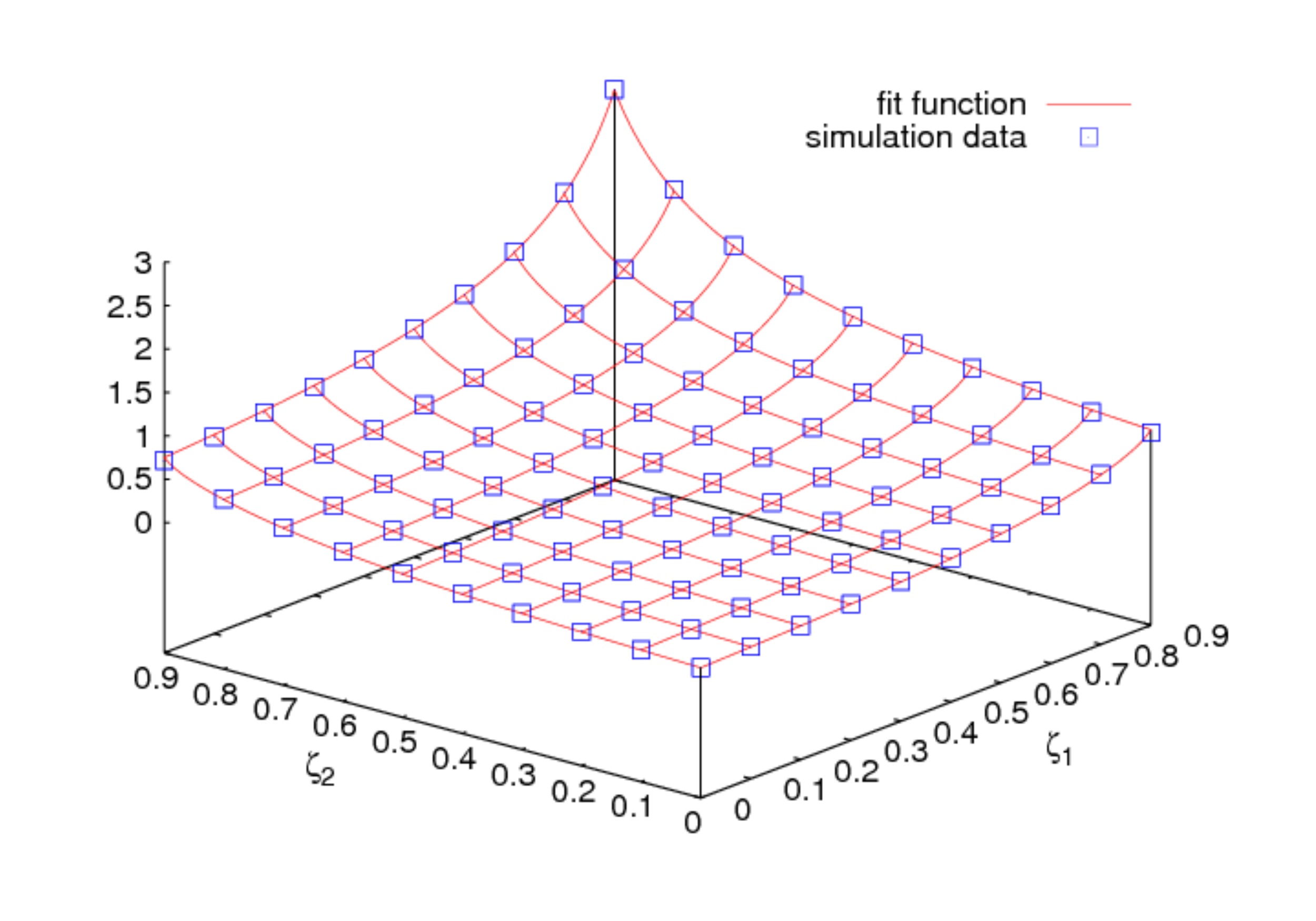}
b)\includegraphics[width=0.47\linewidth]{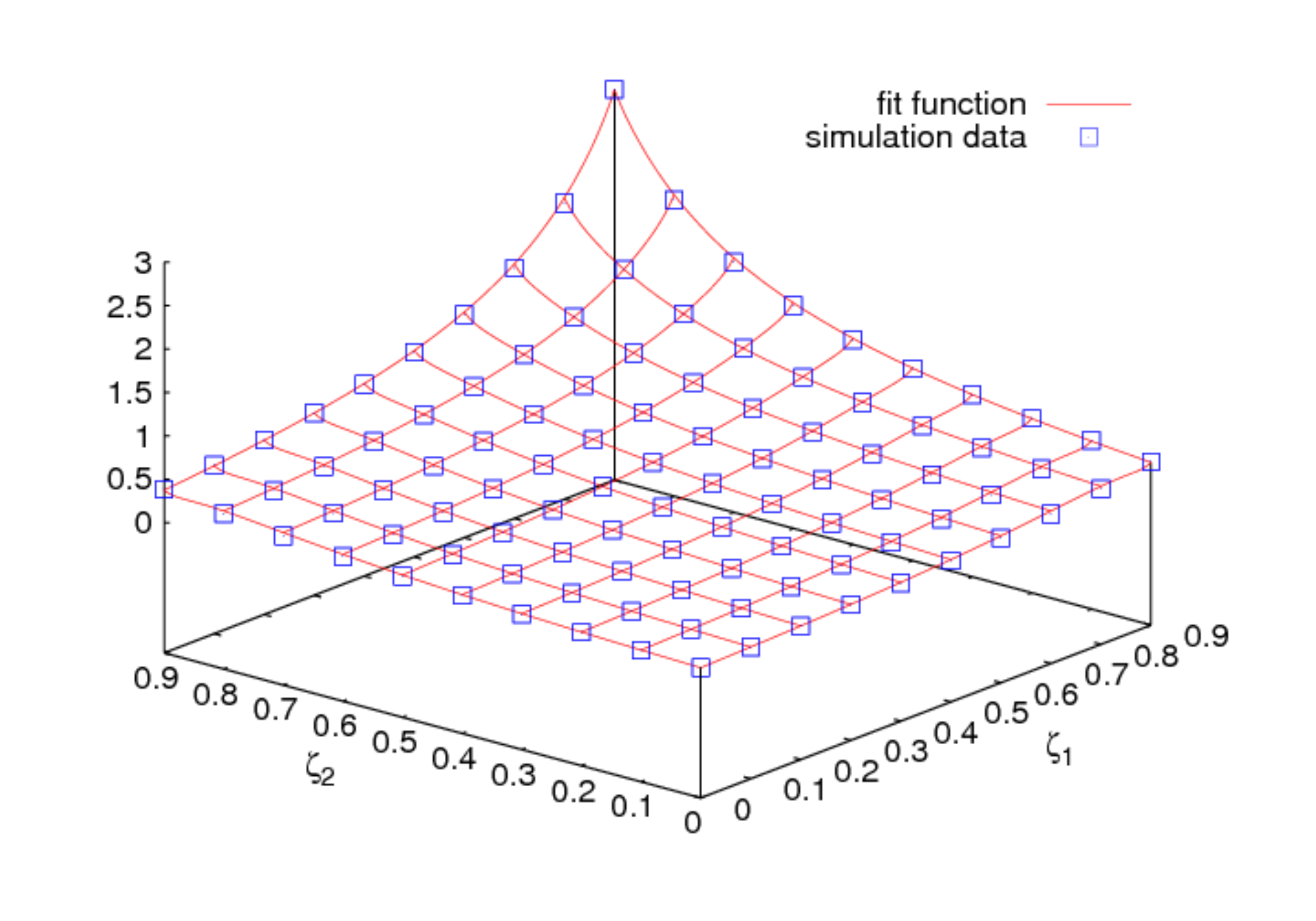}}
\end{minipage}
\caption{Dependence of the effective slip length on the slip parameter of the stripes, according to the stripe orientation being a) parallel and b) perpendicular to the direction of the accelerating force causing the flow. This figure shows that \eq{slip_mod} gives a nearly perfect fit to the simulation data obtained as long as the local slip length is small compared to the width of the stripes.}
\label{fig:stripes_3dfit}
\end{figure*}

Throughout our simulations, we use stripes of 4 lattice units as width, and we observe a perfect third order fit for \eq{slip_mod} as seen in \fig{fig:stripes_3dfit}.

For the fit parameters $A$, $B$, $C$, $D$, and $E$ in \eq{slip_mod}, depending on the orientation of the stripes to the direction of the accelerating force causing the flow, we find the values listed in \tab{tab:parameters}. 

\begin{table}
\begin{minipage}{\linewidth}
\mbox{
\begin{tabular}{|l||r|r|r|r|r|}
\hline
Orientation & $A$ & $B$ & $C$ & $D$ & $E$ \\\hline\hline
$Parallel $ & $0.5$ & $-0.15$ & $0.3$ & $-0.02223$ & $-0.02223$\\\hline
$Perpendicular$ & $0.5$ & $-0.25 $ & $0.5$ & $-0.04167$ & $0.04167$\\\hline
\end{tabular}
}
\end{minipage}
\caption{Fit parameters $A$, $B$, $C$, $D$, and $E$ in \eq{slip_mod} in the case of 
the stripes oriented parallel to the force, and perpendicular to the driving force.} 
\label{tab:parameters}
\end{table}

The flow is driven by an accelerating force which is always applied in the whole domain. Consistent with the work by Feuillebois\etal\,\cite{Feuillebois}, we find 
that the maximum slip length is obtained when the stripes are aligned parallel to this force. 


The dependence of the slip length for walls with stripes oriented at the angle $\theta$ between the 
direction of the force and the extension of the stripes has been 
previously calculated analytically by Bazant and Vinogradova\,\cite{Bazant}, to be
\begin{equation}
  b = b_\parallel \cos^2\theta + b_\perp \sin^2 \theta\,.\label{eq:bazant}
\end{equation}
We find that the slip length obtained in our simulations using the tilted force agrees with 
the theoretical prediction of \eq{eq:bazant} with a relative error of only $10^{-3}\%$. 
\begin{figure}[h]
\centering
\includegraphics[width=\figwidth]{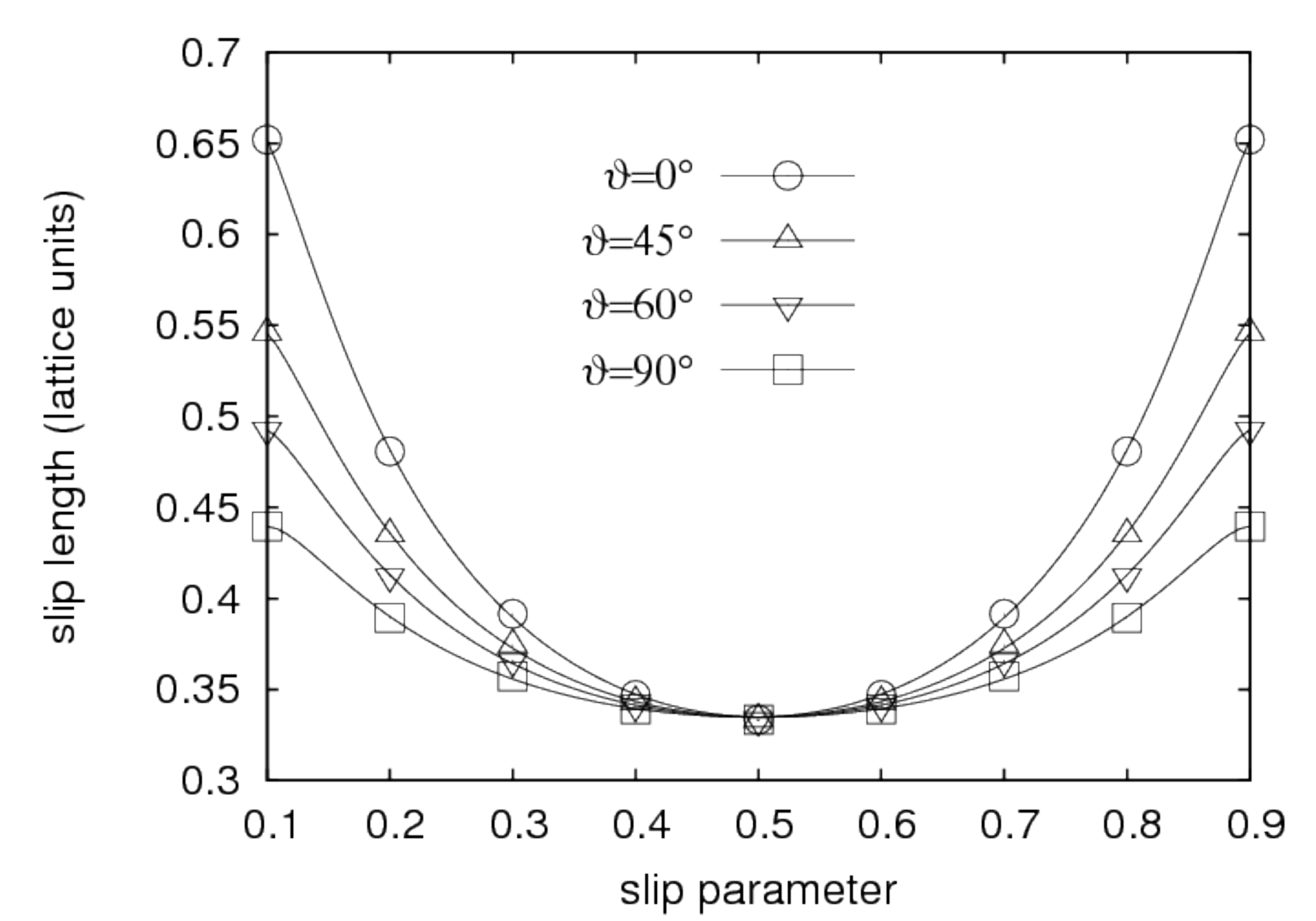}
\caption{Slip length of a striped wall as a function of the slip parameter of the stripes 
for four different orientations of the stripes with respect to the direction of the 
driving force: parallel ($\theta = 0^\circ$), perpendicular ($\theta = 90^\circ$), 
diagonal ($\theta = 45^\circ$), and tilted by 60 degrees. The condition $\zeta_1 +\zeta_2 = 1$ is maintained, leading to symmetry about $\zeta=0.5$.
%
}
\label{fig:sliplength_diag_fit}
\end{figure}
In \fig{fig:sliplength_diag_fit} we show the dependence of the effective slip length on the orientation of the force and on the choice of slip parameters applied on the stripes. For studying the angle dependence, the stripes have been constrained by the condition $\zeta_1 + \zeta_2 = 1$. 
For $\zeta_1 = \zeta_2 = 0.5$, the case of a homogeneous wall is restored and therefore, the slip length is independent on the angle $\theta$ between the force and the direction of the extension of the stripes. If $\zeta_1$ and $\zeta_2$ are chosen differently, the effective slip depends on the orientation. 
For $\theta=0^\circ$, i.e. if the force is aligned parallel to 
the stripes, a maximal slip length is obtained. If the stripes are oriented perpendicular to the direction of the force, a minimal slip length is obtained. Out of those extreme cases the slip length for the orientations in between can be calculated from \eq{eq:bazant}.

We also find that in simulations with the force acting in diagonal direction, 
the fluid tends to follow the stripes on the wall. The velocity of the fluid shows
a small component which is perpendicular to the force. This component can be understood
as a consequence of the tensorial nature of the slip as proposed by Bazant and Vinogradova\,\cite{Bazant}. In our simulations, the transverse flux is four orders of magnitude 
less than the velocity obtained in the direction of the force. Even though the effect is small, 
these simulations confirm that walls with stripes of varying slip length can be used in mixing 
devices to cause vortices in micro and nano flows.

\subsection{Continuously varying Striped Walls}
In this section, the slip parameter $\zeta$ is made to vary as a continuous periodic function along the wall nodes in the $y$ and $z$ directions, in the following manner:
\begin{equation}
\zeta = \overline\zeta  + \hat\zeta\cdot \cos (\vec{k}\cdot\vec{x})\,.\label{eq:continuous}
\end{equation}
where the wave vector $\vec{k}$ defines the frequency and direction of the variation of slip parameter $\zeta$. We use an amplitude $\hat\zeta=0.5$ and a mean slip parameter of $\overline\zeta = 0.5$. 
\[
  \lambda = \frac{2\pi}{\left|\vec{k}\right|}\,.
\]
\begin{figure}[h]
\centering
\includegraphics[width=\figwidth]{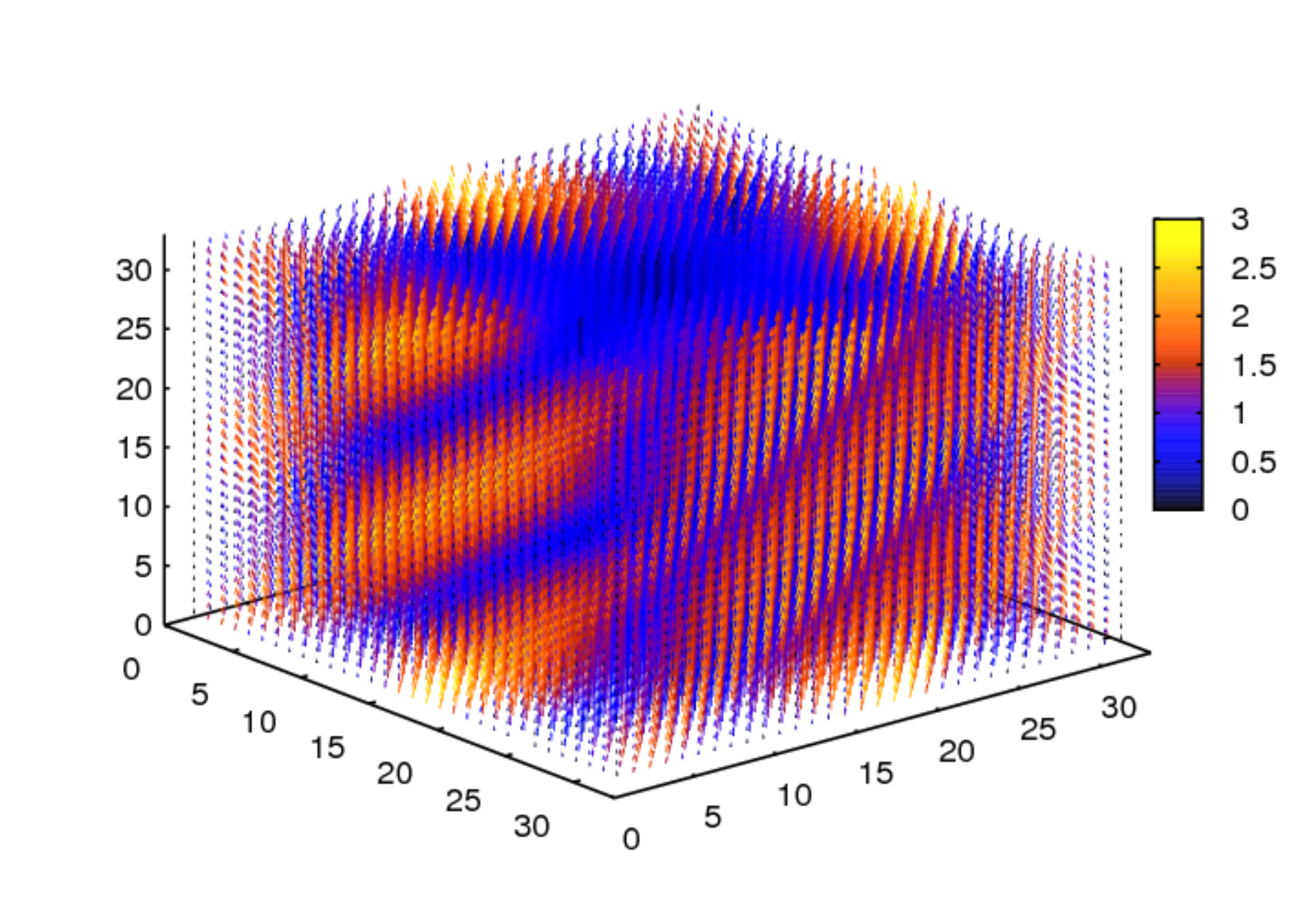}
\caption{Walls are designed having stripes with continuously varying slip parameter $\zeta$, as per \eq{eq:continuous}.
The colors correspond to the velocity projected in the $xy$-plane scaled by a factor of 3000.}
\label{fig:wall3d}
\end{figure}

We run simulations having periodicity of the slip parameter as $\lambda=\frac{31}{2}\sqrt{2}$ lattice nodes, giving rise to stripes as shown in \fig{fig:wall3d}. The wave vector points in the diagonal direction on each of the boundary planes and the magnitude is chosen such that along each side two stripes fit between the edge nodes.

\begin{figure}[h]
\centering
\includegraphics[width=\figwidth]{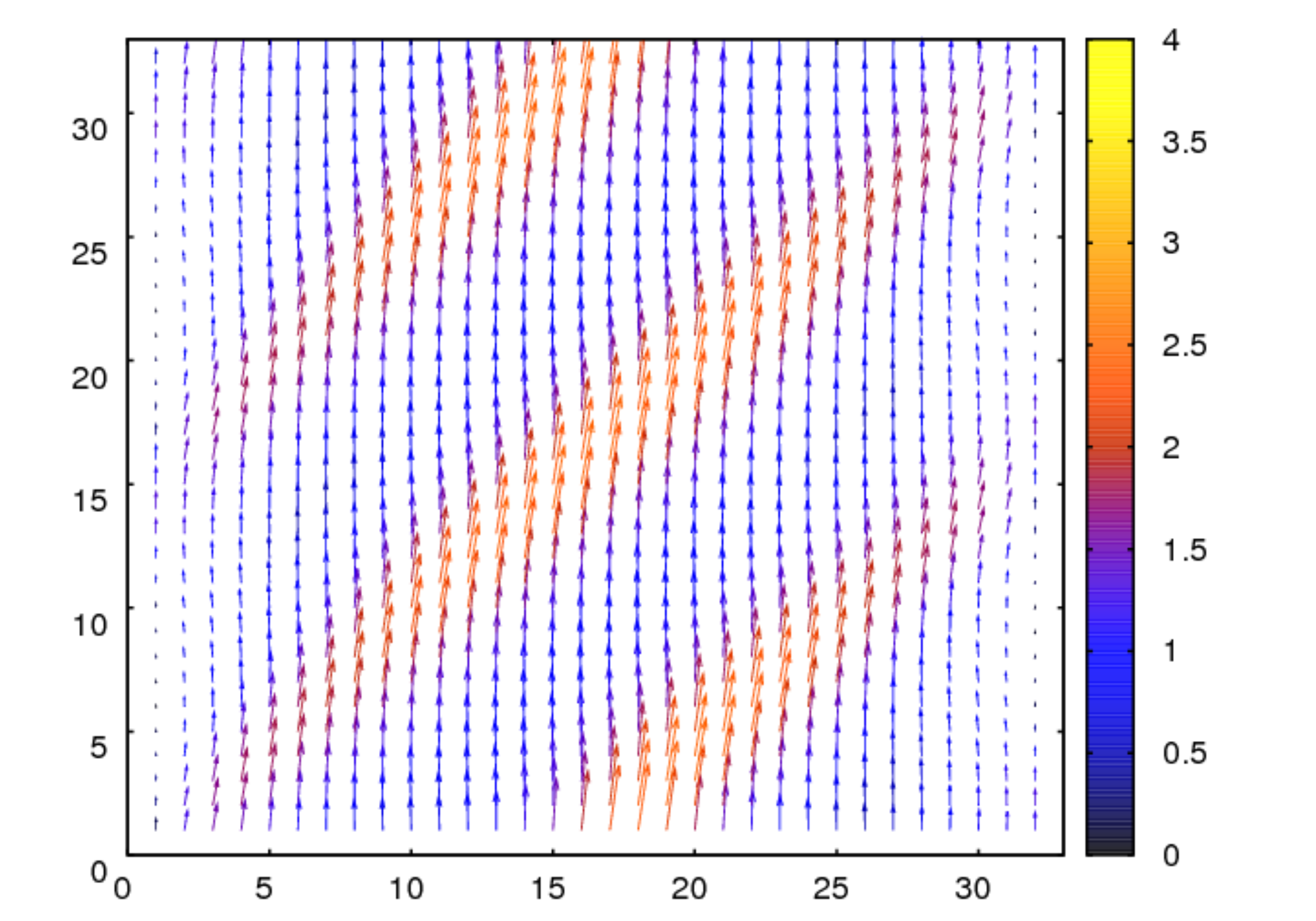}
\caption{The velocity along the $x$ boundary in the case of variation in the slip parameter along the wall nodes. The transverse velocity component is color coded after scaling by a factor of 3000.}
\label{fig:velocitysideview}
\end{figure}

\begin{figure}[h]
\centering
\includegraphics[width=\figwidth]{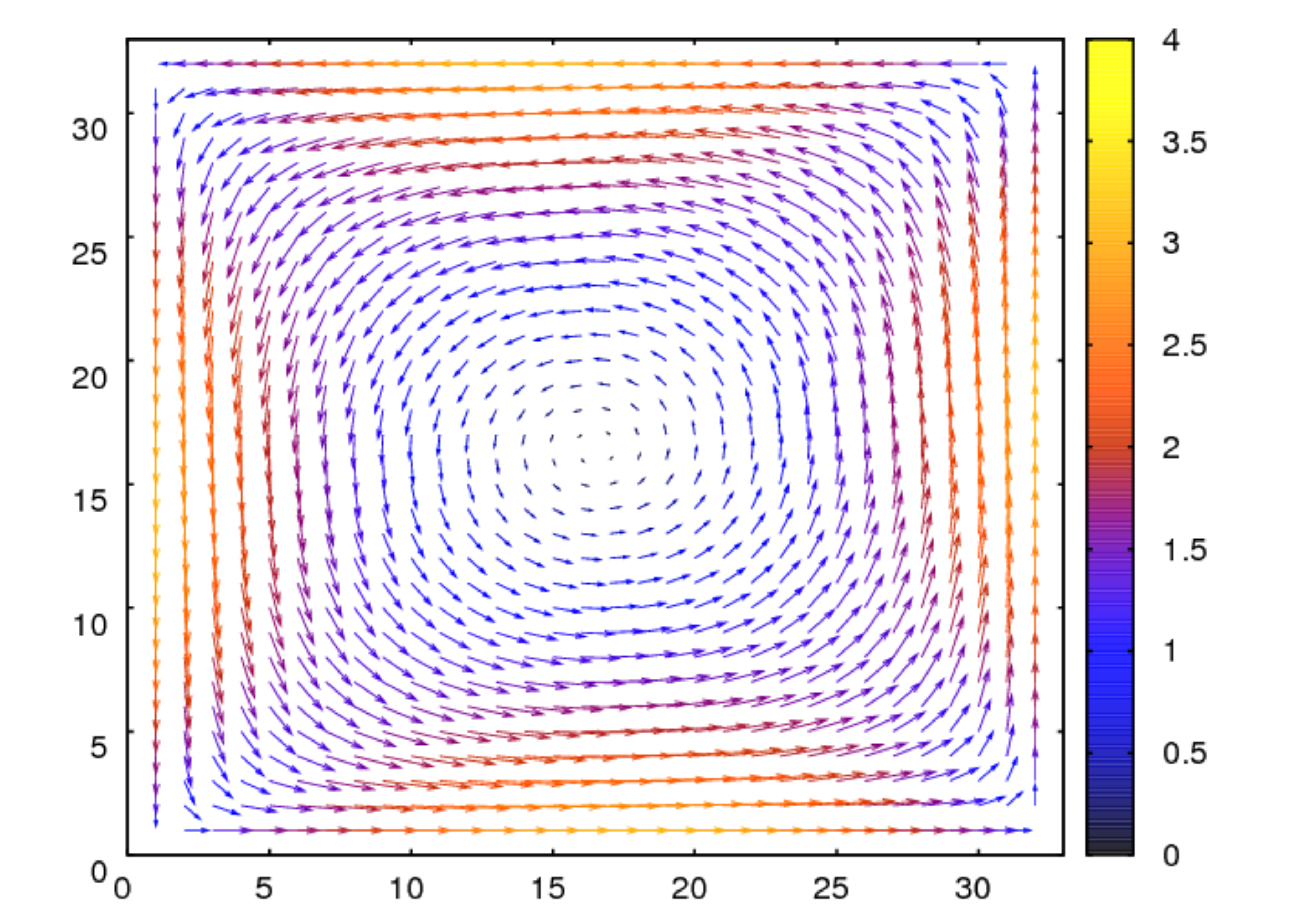}
\caption{An averaged projection of velocity vectors of the $xy$-plane in the case of variation in the slip parameter along the wall nodes, along the entire $z$-direction domain. It shows the existence of a homogeneous rotation about the centre. The velocity at the centre is aligned with the channel, whereas everywhere else a net circular is seen.}
\label{fig:averaged}
\end{figure}

The influence of such a wall design can be understood by plotting the velocity vectors along the wall, as shown in \fig{fig:velocitysideview}. It is clearly seen that as $\zeta$ tends to 1, the velocity along such nodes is increased, and as $\zeta$ tends to 0, the velocity along such nodes is reduced. As a result the fluid arriving in a region where the slip length decreases has to go aside. Hence the occurrence of a vortice in the plane perpendicular to the direction of the force driven flow may be observed. Due to the varying slip conditions a varying direction of the velocity occurs along the $z$ direction. However, if averaged over all lattice planes a net flux in a vortex remains, as shown in \fig{fig:averaged}.

\begin{figure}[t]
\centering
\includegraphics[width=\figwidth]{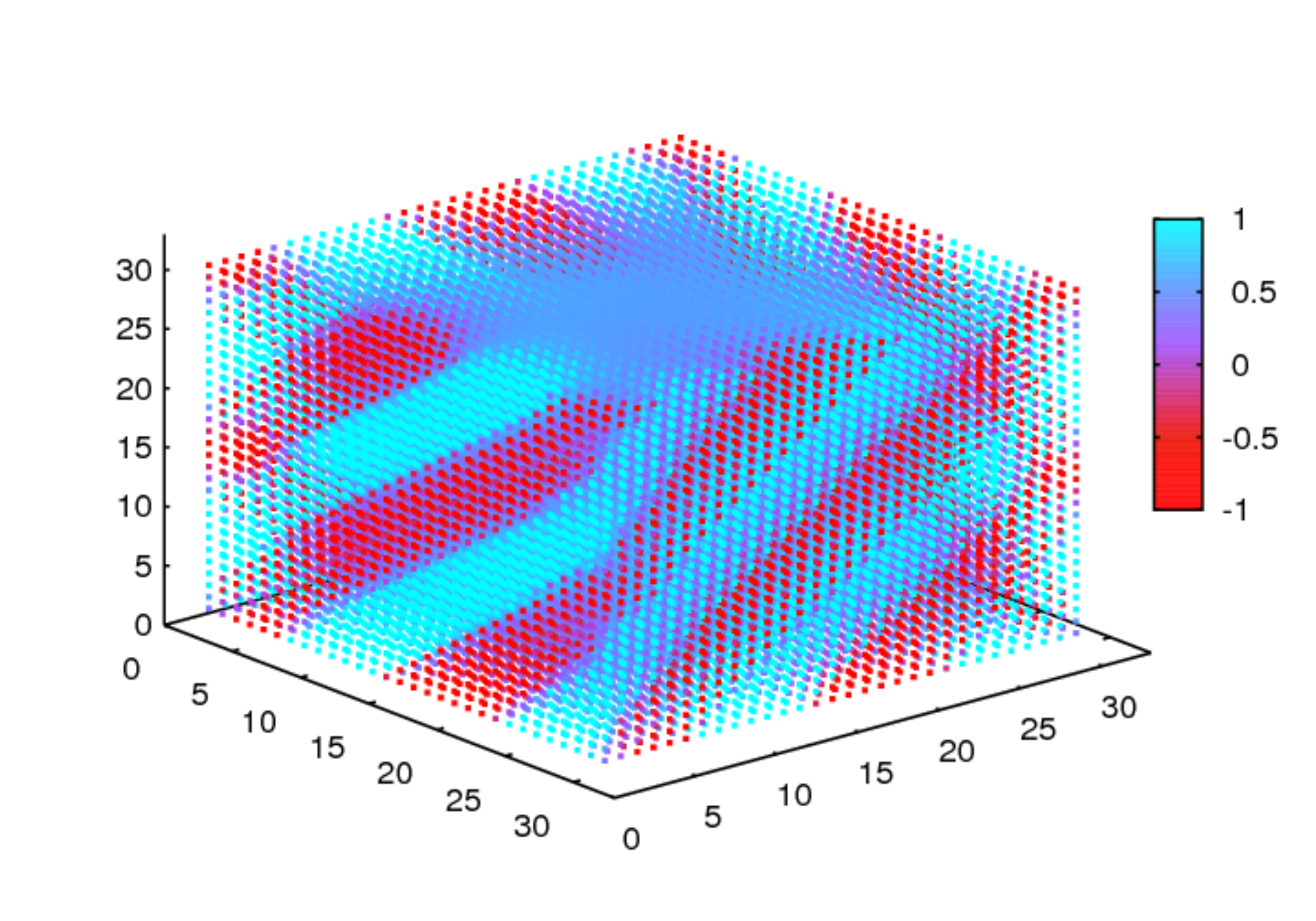}
\caption{A 3D plot of the z-component of the curl of the velocity field in the case of variation in $\zeta$ along the wall nodes, shows the distortions of the flow field by the pattern on the wall and reflects the homogeneous rotation about the centre again. The value calculated for the z-component
of the curl is scaled by a factor of $10^4$ for plotting.}
\label{fig:curl3d}
\end{figure}

In order to better quantify the influence of the pattern on the velocity profile, we calculate the $z$-component of the curl of the velocity field. As shown in \fig{fig:curl3d}, close to the surface, there are distortions by the pattern, but in the center of the channel a homogeneous rotation of the velocity field about the axis of the channel is observed. 

\begin{figure}[h]
\centering
\includegraphics[width=\figwidth]{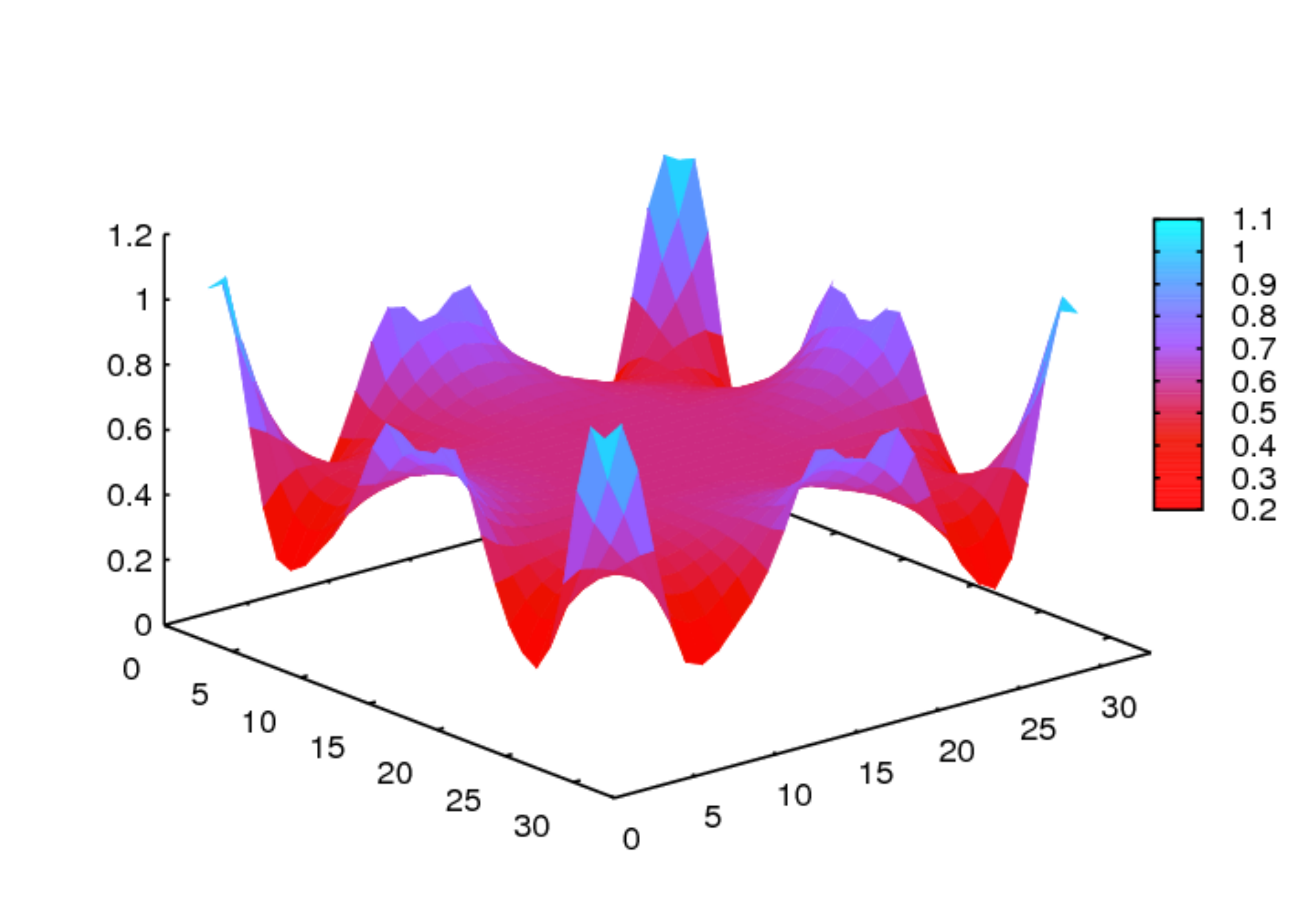}
\caption{The $z$-component of the curl of the velocity field averaged along the $z$-direction in the case of variation in $\zeta$ along the wall nodes. Here the homogeneous rotation at the center of the channel is reflected by the constant value of the curl. The value calculated for the z-component
of the curl is scaled by a factor of $10^4$ for plotting.}
\label{fig:curlprojection}
\end{figure}

If the $z$-component of the curl is averaged along the $z$-direction, indeed a net rotation which is constant in large areas of the cross section is identified. In \fig{fig:curlprojection} this is plotted for both cases of small and large periodicity of the pattern. One can see that with the small period of the pattern the rotation is more homogeneous, whereas with a large period the distortion of the flow is stronger, and the curl obtained in the center is slightly enhanced.

Our results confirm previous understanding that at a decreased angle (stripes along the driving force) the slip length increases. We further understand that if the stripes are oriented parallel to the force, the fluid can flow along the stripes and form lamellae of fluid with different slip according to the local surface properties on the respective stripe.

\section{Conclusion}
In this paper we have applied the boundary condition of Ref.\,\cite{Ahmed09a} to patterned surfaces. We have studied the dependence of the effective slip length in channels with textured surfaces of stripes of equal width, with alternating local slip length. We find that the effective slip length can be expressed as function of the individual slip lengths on the stripes. 


The tensorial nature of the slip can be exploited to make the fluid flow follow
a given direction. We have tested this in simulations where the flow through a 
rectangular channel patterned with diagonal stripes is simulated. The flow follows
the stripes and so a vortex is induced in the flow field. This can be used to
construct a micro mixer device. Our LB simulations can help to
optimize the patterns to put on the surfaces and to study the dependence of 
the flow field on various parameters, such as the period of the pattern or the 
wetting contrast between two stripes. 

\vspace*{2cm}

\section*{Acknowledgment}
N.K.A. thanks the German Academic Exchange Service (DAAD) for a scholarship for the summer of 2009. M.H. thanks the German Research Foundation (DFG) for financial support within grant EAMatWerk. The authors thank Jens Harting for fruitful discussions.


\begin{thebibliography}{10}
\providecommand{\url}[1]{#1}
\csname url@samestyle\endcsname
\providecommand{\newblock}{\relax}
\providecommand{\bibinfo}[2]{#2}
\providecommand{\BIBentrySTDinterwordspacing}{\spaceskip=0pt\relax}
\providecommand{\BIBentryALTinterwordstretchfactor}{4}
\providecommand{\BIBentryALTinterwordspacing}{\spaceskip=\fontdimen2\font plus
\BIBentryALTinterwordstretchfactor\fontdimen3\font minus
  \fontdimen4\font\relax}
\providecommand{\BIBforeignlanguage}[2]{{%
\expandafter\ifx\csname l@#1\endcsname\relax
\typeout{** WARNING: IEEEtran.bst: No hyphenation pattern has been}%
\typeout{** loaded for the language `#1'. Using the pattern for}%
\typeout{** the default language instead.}%
\else
\language=\csname l@#1\endcsname
\fi
#2}}
\providecommand{\BIBdecl}{\relax}
\BIBdecl

\bibitem{Beskok05}
G.~Karniadakis, A.~Beskok, and N.~Aluru, \emph{Microflows and Nanoflows:
  Fundamentals and Simulation}.\hskip 1em plus 0.5em minus 0.4em\relax Springer
  Verlag NY, Heidelberg, 2005.

\bibitem{Rapaport95}
D.~C. Rapaport, \emph{The Art of Molecular Dynamics Simulation}.\hskip 1em plus
  0.5em minus 0.4em\relax Cambridge University Press, 1995.

\bibitem{lauga-brenner-stone}
E.~Lauga, M.~P. Brenner, and H.~A. Stone, ``Microfluidics: The no-slip boundary
  condition,'' in \emph{Handbook of Experimental Fluid Dynamics}, A.~Y.
  C.~Tropea and J.~Foss, Eds.\hskip 1em plus 0.5em minus 0.4em\relax Springer,
  New-York, 2007, ch.~19, pp. 1219--1240.

\bibitem{chen-doolen98}
S.~Chen and G.~D. Doolen, ``Lattice {B}oltzmann method for fluid flow,''
  \emph{Ann. Rev. Fluid Mech.}, vol.~30, p. 329, 1998.

\bibitem{chen-chen-matthaeus}
H.~Chen, S.~Chen, and W.~H. Matthaeus, ``Recovery of the {N}avier-{S}tokes
  equations using a lattice-gas {B}oltzmann method,'' \emph{Phys. Rev.~A},
  vol.~45, no.~8, p. R5339, 1992.

\bibitem{Higuera89}
F.~J. Higuera, S.~Succi, and R.~Benzi, ``Lattice gas dynamics with enhanced
  collisions,'' \emph{Europhys. Lett.}, vol.~9, pp. 345--349, 1989.

\bibitem{Ahmed09a}
N.~K. Ahmed and M.~Hecht, ``A boundary condition with adjustable slip length
  for lattice boltzmann simulations,'' \emph{J. Stat. Mech.}, p. P09017, 2009.

\bibitem{BGK}
P.~L. Bhatnagar, E.~P. Gross, and M.~Krook, ``Model for collision processes in
  gases. {I}. {s}mall amplitude processes in charged and neutral one-component
  systems,'' \emph{Phys. Rev.}, vol.~94, no.~3, p. 511, 1954.

\bibitem{Succi01}
S.~Succi, \emph{The lattice {B}oltzmann equation for fluid dynamics and
  beyond}.\hskip 1em plus 0.5em minus 0.4em\relax Oxford University Press,
  2001.

\bibitem{Hecht2009}
M.~Hecht and J.~Harting, ``General on-site velocity boundary conditions for
  lattice boltzmann,'' \emph{arXiv:0811.4593}, 2008.

\bibitem{Feuillebois}
F.~Feuillebois, M.~Z. Bazant, and O.~I. Vinogradova, ``Effective slip over
  superhydrophobic surfaces in thin channels,'' \emph{Phys. Rev. Lett.}, no.
  102, p. 026001, 2009.

\bibitem{Bazant}
M.~Z. Bazant and O.~I. Vinogradova, ``Tensorial hydrodynamic slip,''
  \emph{J.~Fluid Mech.}, vol. 613, pp. 125--134, 2008.

\end{thebibliography}

\end{document}